\documentclass[twocolumn,prl,twoside,a4paper]{revtex4}
\usepackage{graphicx}
\usepackage{amssymb}

\begin{document}

\title{ Imaging correlations in heterodyne spectra for quantum displacement sensing}
\author{A.~Pontin}
\address{Department of Physics and Astronomy, University College London, Gower Street, London WC1E 6BT, United Kingdom}
\author{J.E.~Lang}
\address{Department of Physics and Astronomy, University College London, Gower Street, London WC1E 6BT, United Kingdom}

\author{A.~Chowdhury}
\address{Istituto Nazionale di Fisica Nucleare (INFN), Sezione di Firenze, Via Sansone 1, I-50019 Sesto Fiorentino (FI), Italy}
\address{CNR-INO, L.go Enrico Fermi 6, I-50125 Firenze, Italy}
\author{P.~Vezio}
\address{Dipartimento di Fisica e Astronomia, Universit\`a di Firenze, Via Sansone 1, I-50019 Sesto Fiorentino (FI), Italy}
\author{F.~Marino}
\address{Istituto Nazionale di Fisica Nucleare (INFN), Sezione di Firenze, Via Sansone 1, I-50019 Sesto Fiorentino (FI), Italy}
\address{CNR-INO, L.go Enrico Fermi 6, I-50125 Firenze, Italy}
\author{B.~Morana}
\address{Institute of Materials for Electronics and Magnetism, Nanoscience-Trento-FBK Division, 38123 Povo (TN), Italy}
\address{Delft University of Technology, Else Kooi Laboratory, 2628 Delft, The Netherlands}
\author{E.~Serra}
\address{Delft University of Technology, Else Kooi Laboratory, 2628 Delft, The Netherlands}
\address{Istituto Nazionale di Fisica Nucleare, TIFPA, 38123 Povo (TN), Italy}
\author{F.~Marin}
\address{Istituto Nazionale di Fisica Nucleare (INFN), Sezione di Firenze, Via Sansone 1, I-50019 Sesto Fiorentino (FI), Italy}
\address{CNR-INO, L.go Enrico Fermi 6, I-50125 Firenze, Italy}
\address{European Laboratory for Non-Linear Spectroscopy (LENS), Via Carrara 1, I-50019 Sesto Fiorentino (FI), Italy}

\author {T.~S.~Monteiro}
\email{t.monteiro@ucl.ac.uk}
\address{Department of Physics and Astronomy, University College London, Gower Street, London WC1E 6BT, United Kingdom}

\begin{abstract}
The extraordinary sensitivity of the output field of an optical cavity to small quantum-scale displacements has led to breakthroughs  such as the first  detection of gravitational waves and of the motions of quantum ground-state cooled mechanical oscillators. 
While heterodyne detection of the output optical field of an optomechanical system exhibits asymmetries which provide a key signature that the mechanical oscillator has attained the quantum regime,  important quantum correlations
are lost. In turn,  homodyning can detect quantum squeezing in an optical quadrature, but loses the important  sideband asymmetries.
 Here we introduce and experimentally demonstrate a new technique, subjecting
the autocorrelators of the output current to filter functions, which restores the lost heterodyne correlations (whether classical or quantum), drastically augmenting the useful information accessible. The filtering even adjusts for moderate errors in the locking phase of the local oscillator. Hence we demonstrate single-shot measurement of hundreds of different field quadratures allowing rapid imaging of detailed features from a simple heterodyne trace. We also obtain a spectrum of hybrid homodyne-heterodyne character,  with motional sidebands of combined amplitudes  comparable to homodyne. Although investigated here in a thermal regime, the method's robustness and generality represents a promising new approach to sensing of quantum-scale displacements.
\end{abstract}

\maketitle

\begin{figure*}[ht]
\begin{center}
{\includegraphics[width=6.in]{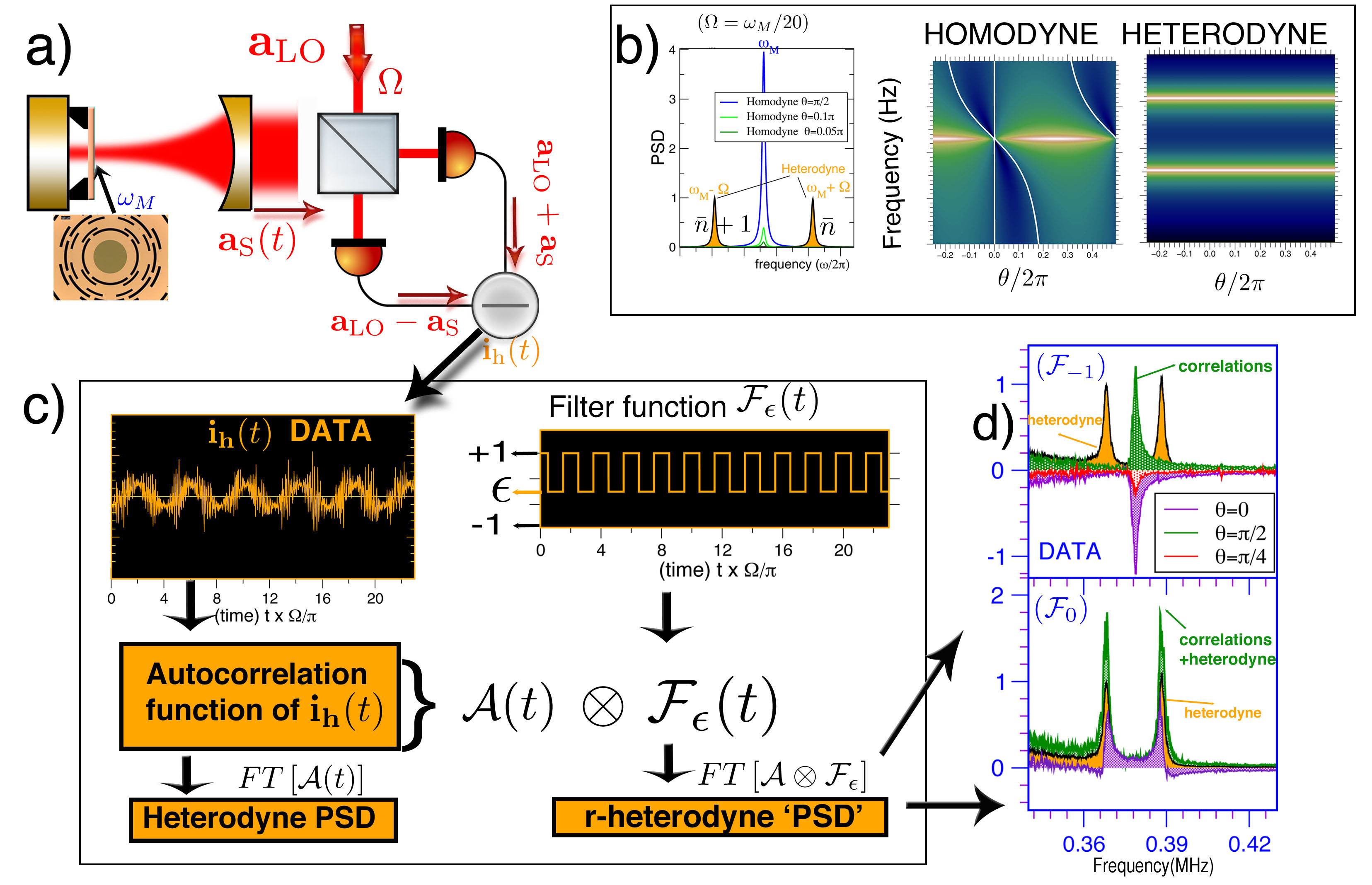}}
\end{center}
\caption{ {\bf (a)} Experimental scheme:  a membrane-in-the-middle set-up within a cavity with a
single optical mode coupled to multiple mechanical modes of the membrane, simultaneously cooling and detecting them. The cavity output signal is beat with a reference oscillator $a_{LO} \propto e^{i\Omega t +\theta}$ of beat frequency $\Omega$ and phase $\theta$. Balanced detection produces an output current $i_h(t)$; then $\Omega=0$ corresponds to homodyne detection while $\Omega\neq 0$ corresponds to heterodyne. 
The output field from an optical cavity provides exquisitely sensitive detection of even quantum-scale displacements via the Power Spectral Density (PSD) of $i_h(t)$. {\bf (b)} Schematic illustration of homodyne vs heterodyne features: while heterodyning  detects quantum regimes via the  dependence of the motional sidebands on phonon occupancy $\bar{n}$, their combined height is only half of the homodyne amplitude. For homodyne detection  the presence of field noise correlations yields squeezing, evidenced by PSD  values below the quantum imprecision noise floor (the region between white lines of the map). In  contrast, the PSD of the heterodyne spectra shows neither squeezing nor any $\theta$ dependence.  {\bf (c)} Outline of the r-heterodyne method relative to the standard heterodyne PSD. {\bf (d)} Example spectra using $\mathcal{F}_{-1}$ (upper panel) and $\mathcal{F}_{0}$ (lower panel).
 Spectra shown normalised to the heterodyne peak amplitudes (orange peaks).
For the upper panel, the correlations are isolated from the main peaks; for the lower panel the restored correlations interfere with the heterodyne PSDs, doubling the  sideband height for $\theta=\pi/2$. }
\label{Fig1}
\end{figure*}

Homodyne and heterodyne detection represent  ``twin-pillars'' of quantum
displacement sensing using the output field of an optical cavity, having permitted major breakthroughs including
detection of gravitational waves \cite{LIGO,LIGODC}. Earlier versions of LIGO employed a radio-frequency (RF) heterodyne detection system, but this was later replaced by a homodyne scheme
\cite{LIGODC}. The broader field of quantum cavity optomechanics has also exposed a rich seam of interesting phenomena
arising from the coupling between the mode of a cavity and a small mechanical oscillator
\cite{Bowenbook,AKMreview,Braginsky} and of the motion of quantum ground-state cooled mechanical oscillators.
Several groups have successfully cooled a mechanical oscillator
\cite{Teufel2011,Chan2011,Kipp2012}  down to  mean phonon occupancy $\overline{n} \sim 1$ or under,
 close to its quantum ground state. Read-out of the temperature was achieved by detection of motional sidebands in the cavity output by homodyne or heterodyne methods.

\begin{figure*}[ht]
\begin{center}
{\includegraphics[width=6.in]{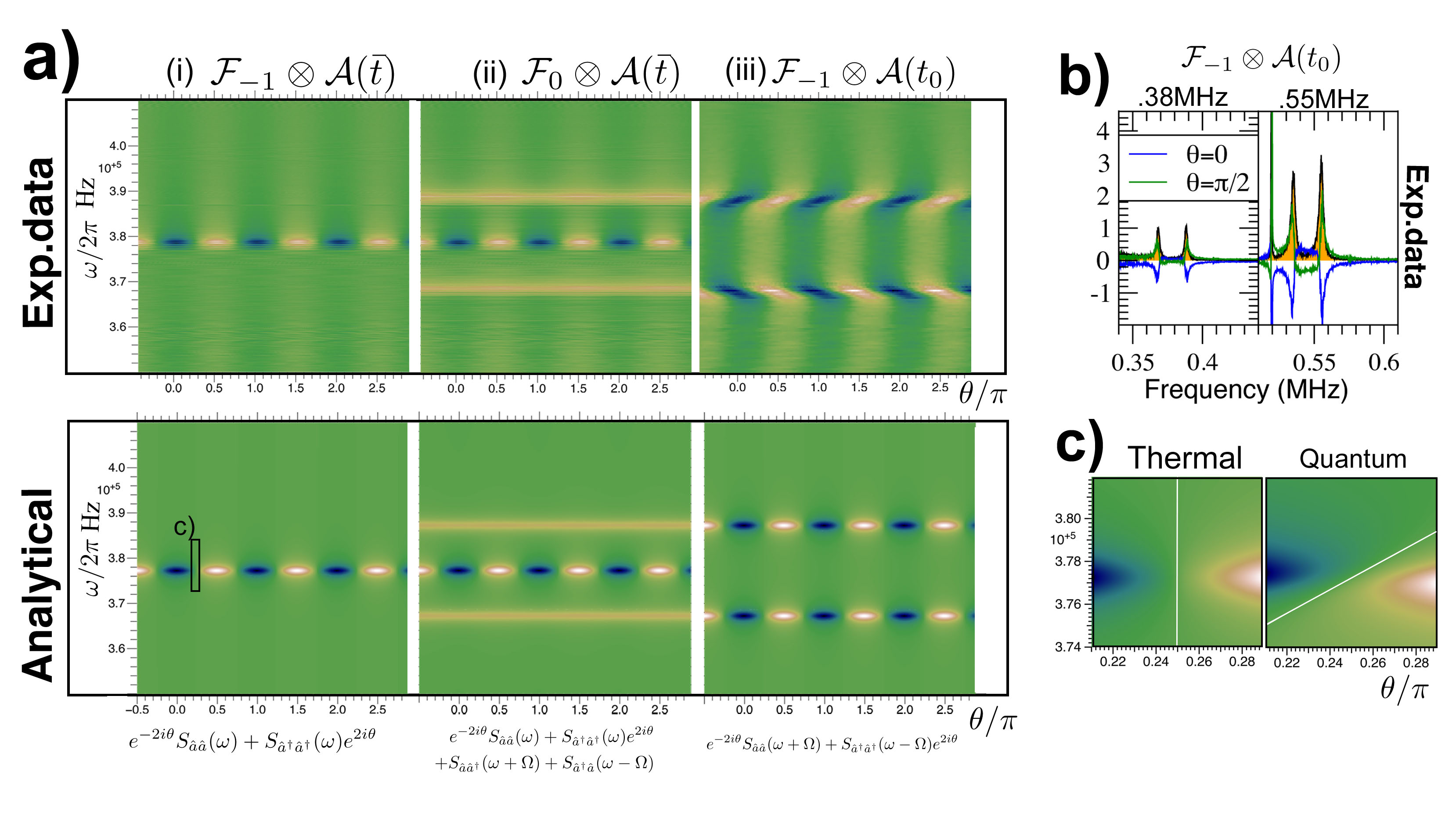}}
\end{center}
\caption{
  {\bf (a)} Mapping correlations: in contrast to the usual featureless ($\theta$-independent) heterodyne spectra, the  correlations are extracted efficiently, even in an experiment where $\theta$
is not actively stabilised. Figure compares r-heterodyne experimental spectra, subjected to different filters, with analytically calculated spectra.
(i) Distils isolated correlations near $\omega=\omega_M$ (ii) Obtains correlations plus heterodyne peaks
(iii) Distils correlations only, but displaced to the position of the heterodyne peaks.
Agreement between experiment and theory is excellent, even if some distortion is evident in (iii), which we attribute to experimental noises (frequency noise, the drift in  $\theta$). Data is in the thermal regime and all spectra are normalised to the heterodyne peak amplitude of the 0.38 MHz mode. Maps are on a linear colour scale where white=1 and dark blue =-1. {\bf (b)}
 Illustrates two spectra i.e. ``cuts'' from (iii) and compares with heterodyne (orange) showing that despite distortion, the strongest $\theta=\pi/2$ peak amplitudes remain close to the expected amplitude
 ($2/\pi$ of the heterodyne amplitude).
 {\bf (c)} Inset of (i) Thermal vs pure quantum back-action spectra (analytical) illustrating
future possible signatures of quantum regimes facilitated by high-resolution mapping of correlations:
  the zero contour (white) rotates away from the vertical as the quantum back-action
 contribution becomes dominant.}
\label{Fig2}
\end{figure*}

Heterodyne (but not homodyne) detection exposes mechanical sideband asymmetries \cite{SideAsymm2012,SideAsymm2012a,Weinstein} which are the hallmark of the quantum regime \cite{SideAsymm2012a,Weinstein}: the observations mirror an underlying asymmetry in the  motional spectrum since an oscillator in its ground state $\overline{n}=0$, can absorb a phonon and down-convert the photon frequency (Stokes process); but it can no longer emit any energy and up-convert a photon (anti-Stokes process). It has also been used to establish  cooling limited by only quantum backaction \cite{Peterson}.
Homodyne detection, on the other hand, allows detection of ponderomotive squeezing, whereby narrowband cavity output field fluctuations fall below the optical shot noise level \cite{Safavi2013,Purdy2013,Pontin}, enhancing quantum measurement sensitivity relative to heterodyne.

 Here we introduce a  method, using filter functions in the usual construction of power spectral densities (PSD) which fully restores the lost quantum correlations to a simple heterodyne measurement,  offering a ``best of both worlds'' scenario in some contexts or at least providing hybrid heterodyne-homodyne spectra retaining some advantages of both methods.  We demonstrate experimentally that one can either detect the correlations in isolation (hence without imprecision noise floor in the frequency spectra) or in interference with the primary peaks; the method is robust to moderate variations in the local oscillator phase. We term the broad approach to retrospective recovery of lost heterodyne correlations ``r-heterodyning''. 

 In Fig.\ref{Fig1}(a), we present schematically the experimental scheme.
The data time series required for our analysis were obtained by operating a cavity optomechanics set-up based on the membrane in the middle configuration. The circular, high-stress SiN$_x$ membrane (shown as photographic inset Fig.\ref{Fig1}(a)) is integrated in an on-chip ``loss shield'' structure \cite{borrielli} and placed in a Fabry-Perot cavity of length $4.38$~mm and cavity Finesse of $\mathcal{F}\approx 13000$ (half-linewidth $\kappa=1.3 \ \textrm{MHz} \times 2\pi$). The membrane was placed at a fixed position, $2$~mm, from the cavity output mirror. The driving beam entering the cavity, from a Nd:YAG 
laser at $1064$~nm, was locked to the cavity by means of the PDH (Pound-Drever-Hall) technique, but was kept detuned from the optical resonance. The input power was  $P_{in}=70~\mu$W, half of which lies in the carrier. The reflected field was divided in two beams, $\sim 30\%$ is used in PDH detection while the remaining $\sim 70\%$ was superimposed on a $500$~$\mu$W local oscillator (LO) field, shifted in frequency by $\frac{\Omega}{2\pi}=10$~kHz with respect to the driving beam. 

The ensuing balanced detection provided the photocurrent time traces that we employed here. The resulting photocurrent time-series, takes the (appropriately normalised) form
$i_h(t)= {\hat{a}}^\dagger e^{i(\Omega t+\theta)}+ {\hat{a}}e^{-i(\Omega t+\theta)}$.
Although the measured current is real, for generality we allow the underlying fields to correspond to quantum fluctuations.

In optomechanics experiments, the interesting dynamical features appear in frequency space as peaks (sidebands) of width $\sim \Gamma$ where $\Gamma$ is total damping, near  $\omega \simeq \pm \omega_M$ where $\omega_M$ is the mechanical frequency. Analysis of such narrowband data features typically proceeds via the PSD (Power Spectral Density) of the measured current, 
$ S_{ i_h i_h}(\omega) \equiv \langle |{\hat i}_h(\omega)|^2\rangle$.

Homodyne detection corresponds to $\Omega=0$ so leads to detection of a single quadrature
$\hat{Y}_\theta={\hat{a}}^\dagger e^{i\theta}+ {\hat{a}}e^{-i\theta}$  and the corresponding symmetrised \cite{homodyne}  PSD
 of the current.  The homodyne PSD  may be written :
\begin{equation}
S_{Y_\theta Y_\theta}(\omega) = S_{\hat{a} \hat{a}^\dagger}(\omega)+S_{\hat{a}^\dagger \hat{a}}(\omega)
+S_{\hat{a}^\dagger \hat{a}^\dagger}(\omega)e^{2i\theta}+ S_{\hat{a} \hat{a}}(\omega)e^{-2i\theta} \nonumber \\
\label{homo}
\end{equation}
%where we define $ S_{{\hat Q}^\dagger {\hat Q}}(\omega) \equiv \langle |{\hat Q}(\omega)|^2\rangle$ for 
% $\hat{Q}\equiv \hat{a}^\dagger \ \textrm{or} \ \hat{a}$.
All the $\theta$-dependence is in the  $S_{\hat{a}\hat{a}}$ and $S_{\hat{a}^\dagger\hat{a}^\dagger}$ terms.
These terms embody the spectral correlations between the amplitude $\hat{Y}_0$ and phase
$\hat{Y}_{\pi/2}$ of the cavity field and where they arise from  back-action from the laser shot-noise, their presence is taken as an important quantum signature
\cite{KippNun2016,Purdy2017,Lehnert2017}.

In contrast, for $\Omega > 0$, heterodyne detection yields the PSD:
\begin{equation}
S^{\Omega}_{het}(\omega)= S_{\hat{a}\hat{a}^\dagger}(\omega+\Omega)
              +S_{\hat{a}^\dagger\hat{a}}(\omega-\Omega)
\label{heterodynePSD}
\end{equation}
since for a time-trace of reasonable duration $T$ where $T \gg 2\pi/\Omega$, the temporal rotation
of the measured  quadrature $\equiv\hat{a} e^{-i(\theta+\Omega t) }+\hat{a}^\dagger e^{i(\theta+\Omega t)}$
leads to averaging out and loss of the $S_{\hat{a}\hat{a}}$
and $S_{\hat{a}^\dagger\hat{a}^\dagger}$ components.

 Fig.\ref{Fig1}(b) illustrates the fact that the combined amplitudes of the mechanical sidebands are only one half of the homodyne case.
The homodyne PSDs have strong dependence on $\theta$ and 
 can drop below the shot noise level (the region between the white lines in the Fig.\ref{Fig1}(b) map), indicating (ponderomotive) squeezing of the optical quadratures. 
 Heterodyne  PSDs are $\theta$-independent but yield two, rather than one, sidebands near $\omega \simeq +\omega_M$. Although $S^{\Omega}_{het}(\omega)=S^{\Omega}_{het}(-\omega)$ the sidebands are not symmetric about
$\omega=\Omega$ in quantum regimes, where such Stokes/anti-Stokes sideband asymmetry also represents a valuable quantum signature \cite{Peterson,KippNun2016}.

The transition between homodyne and heterodyne is too abrupt to easily 
probe.  Since  $T   \gg \Gamma^{-1} \gg  2\pi/\Omega$ the heterodyne sidebands of Fig.\ref{Fig1}(b) become unresolvable as $\Omega \to 0$ before any correlations of the true $\Omega=0$ limit are manifest in the spectrum.
Hence one cannot normally measure PSDs ``intermediate'' between  heterodyne and homodyne.

\begin{figure}[ht]
\begin{center}
{\includegraphics[width=2.1in]{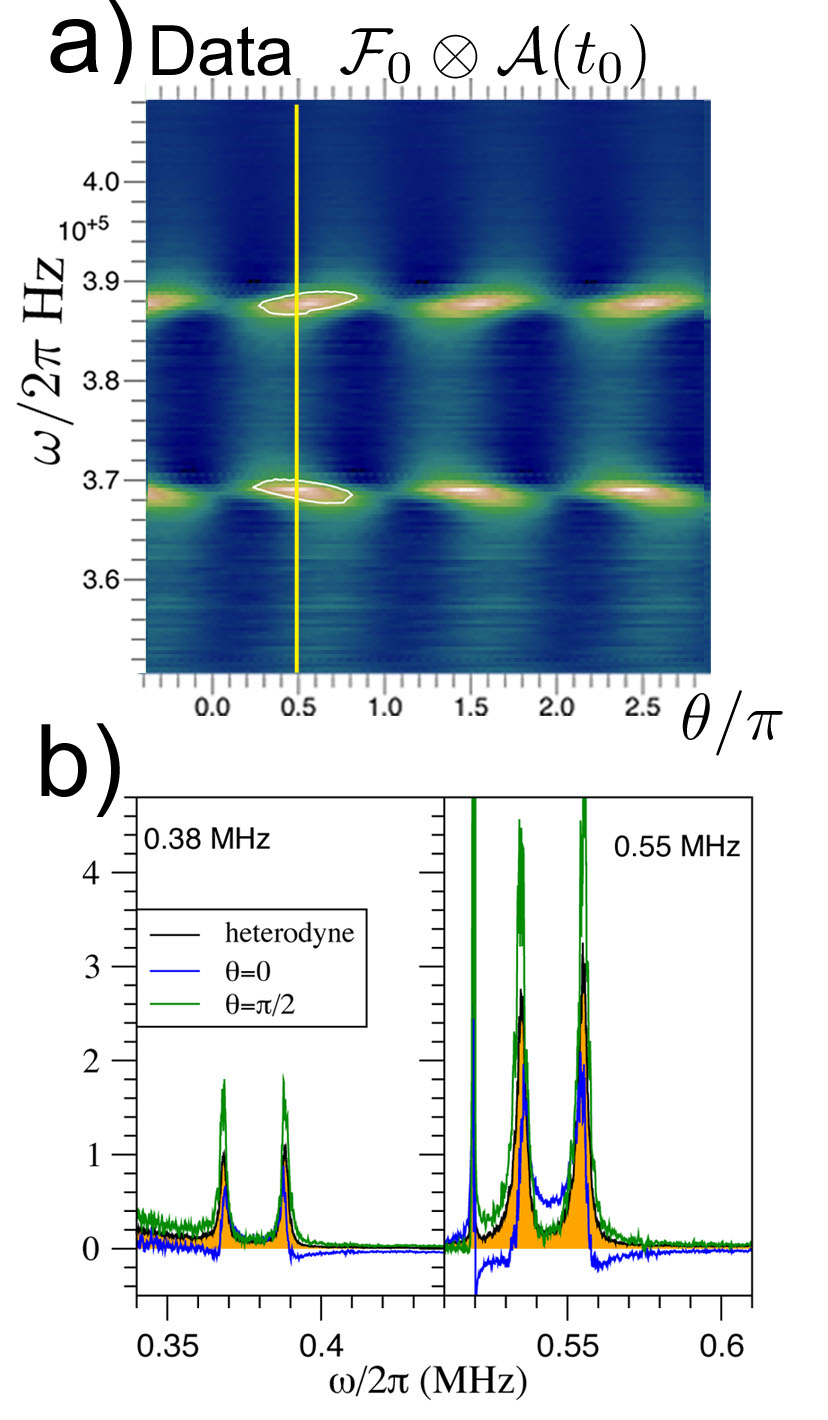}}
\end{center}
\caption{
  {\bf (a)} A spectrum of hybrid homodyne-heterodyne character. Map for  $\mathcal{F}_0 \otimes \mathcal{A}(t_0)$ filter that most closely mimics homodyne behaviour, since the
 correlations interfere with the static heterodyne peaks. 
In future experiments, mapping will allow efficient one-shot detection of quantum behavior in the mechanical oscillator (via sideband asymmetry) and in the cavity mode (via quantum squeezing).   
{\bf (b)} Data for two mechanical modes: the r-heterodyne peaks have nearly double the heterodyne amplitudes, hence their combined amplitudes become comparable to the corresponding homodyne values
allowing sensitive displacement detection(white cut in (b) indicates maxima $\theta=\pi/2$). Peaks normalised to heterodyne amplitude of .38 MHz mode. White contour indicates where the heterodyne amplitude is exceeded. Map colours on a linear scale ranging from white$=2$ to dark blue$=0$.}
\label{Fig3}
\end{figure}

However, we show there is a  straightforward
but robust procedure to recover the lost correlations.
This is illustrated in Fig.\ref{Fig1}(c).
 Our departure point is to approach the PSD via the autocorrelator
(rather than from the FT of the time trace). Then there are two choices
for the construction of the autocorrelator:
either (i) $ \mathcal{A}_{t_0}(t)=\frac{1}{T}\int_0^T dt_0 \langle i(t_0)i(t_0+t) \rangle$ 
or (ii) $ \mathcal{A}_{\bar{t}}(t)=\frac{1}{T}\int_0^T d\bar{t}\langle i(\bar{t}-t/2)i(\bar{t}+t/2)\rangle$.
For an ordinary PSD, the above two choices are quite equivalent 
amounting only to a trivial change in coordinates so:
\begin{equation}
S_{ i_h i_h}(\omega)=\textrm{FT}[\mathcal{A}_{t_0}(t)]\equiv\textrm{FT}[ \mathcal{A}_{\bar{t}}(t)]
\end{equation}
making use of the well-known Wiener-Kinchin theorem. We use the convenient shorthand $ \mathcal{A}(t_0) \equiv \mathcal{A}_{t_0}(t)$ in figure labels to label the corresponding time-series
hence $S_{ i_h i_h}(\omega) \equiv \textrm{FT}[ \mathcal{A}(t_0)]$, for example.

Our method is outlined in Fig.\ref{Fig1}(c). Spectra are instead obtained from a
 (non-standard) convolution involving the autocorrelator and a filter function.
We assume long $T$ and modest
$ \Omega \sim \omega_M/10-\omega_M/100 \gg \Gamma$.
Defining $\mathcal{F} \otimes \mathcal{A}_{t_0}(t) \equiv \frac{1}{T}\int_0^T dt_0  \mathcal{F}(t_0)\langle i_h(t_0)i_h(t_0+t) \rangle $, we then write $S^{t_0}_{ i_h i_h}(\omega) \equiv \textrm{FT}[ \mathcal{F} \otimes \mathcal{A}(t_0)]$ or
the  alternative form $S^{\bar{t}}_{ i_h i_h}(\omega) \equiv \textrm{FT}[ \mathcal{F} \otimes \mathcal{A}(\bar{t})]$. Above, $\mathcal{F}$ is a filter function which is resonant with $2\Omega$, but otherwise is chosen for convenience. We define its Fourier transform $\textrm{FT}[\mathcal{F}(t)]\equiv \tilde{\mathcal{F}}(\omega)$ (see \cite{Suppinfo} for discussion of FT limits)
and restrict ourselves to the case $\tilde{\mathcal{F}}(\omega)=\tilde{\mathcal{F}}(-\omega)$.
Here we take $ \mathcal{F}_\epsilon(t)=+1$ for  $2\Omega t =2\pi N +\phi$ for $N=0,1,2...$ and with $\phi\in[-\pi/2:\pi/2]$ and $\mathcal{F}_\epsilon=\epsilon$ elsewhere. We then obtain (see \cite{Suppinfo}):
\begin{equation}
T S_{i_h i_h}^{\bar{t}}(\omega)\simeq \tilde{\mathcal{F}}_\epsilon(\omega=0)S^{\Omega}_{het}(\omega)+ \tilde{\mathcal{F}}_\epsilon(\omega=2\Omega)S_{corr}(\omega).
\label{rheterodyne}
\end{equation}
where $S^{\Omega}_{het}(\omega)$ is given by Eq.\ref{heterodynePSD} and the second term
 $S_{corr}(\omega)=S_{\hat{a}^\dagger \hat{a}^\dagger}(\omega)e^{2i\theta}+ S_{\hat{a} \hat{a}}(\omega)e^{-2i\theta}$
represents the correlations, but {\em at the homodyne position}, hence shifted away from
the usual heterodyne PSD components at $\Omega \pm \omega$.

In contrast, for the $t_0$ case where we obtain spectra using $\textrm{FT}[ \mathcal{F} \otimes \mathcal{A}(t_0)]$, we have $T S_{i_h i_h}^{t_0}(\omega)\simeq \tilde{\mathcal{F}}_\epsilon(0)S^{\Omega}_{het}(\omega)+ \tilde{\mathcal{F}}_\epsilon(2\Omega)S_{corr}(\omega)$ but now :
\begin{equation}
S_{corr}(\omega)=S_{\hat{a}\hat{a}}(\omega-\Omega)e^{-2i\theta}+ S_{\hat{a}^\dagger \hat{a}^\dagger}(\omega+\Omega)e^{2i\theta}
\label{rhett0}
\end{equation}
and the frequencies of the correlation spectra now coincides with that of the usual heterodyne sidebands.

From the above we have enormous flexibility in tuning the relative amplitudes (as well as relative
positions) of the correlations vs the heterodyne PSD by choosing $\epsilon$. Clearly, for $ \epsilon=+1$, an ordinary heterodyne PSD is recovered; for $ \epsilon=-1$, the stationary heterodyne PSD components are completely eliminated. For intermediate values a tunable mixture is obtained.

In this approach, the rotating quadrature heterodyne current is turned to 
advantage as $\theta$-values can be rapidly adjusted by shifting the filter,
allowing single-shot mapping of correlations.
The only restriction is that the heterodyne beat 
frequency note must be modest $\Omega \sim \omega_M/10- \omega_M/100$ but this does not present extra experimental challenges.
Fig.\ref{Fig2}(a) shows the cases $\epsilon=-1$ and $\epsilon=0$ 
where maps are generated for 800 spectra at different $\theta$ and are compared with standard analytical expressions for the quantum noise spectra \cite{Bowenbook,AKMreview} showing excellent agreement.

Experimental data was obtained and analysed  for two different membrane modes: the (1,1) mode with effective resonance, linewidth and mass of, respectively, $378.16$~kHz, $4.56$~kHz and $300$~ng; 
the (0,2) mode with $544.78$~kHz effective resonance, a $8.44$~kHz linewidth and an effective mass of $180$~ng.
As shown in Fig.\ref{Fig2}(b), the behaviour for the different modes is very similar.

The LO phase $\theta$ in our experiment was not actively stabilized. However, the LO phase time series 
$\theta(t)$  was recovered in post-processing by implementing a numerical lock-in amplifier to demodulate the $10$~kHz beat note. In particular, a slow (relative to $\Omega^{-1}$) 
 modulation at $\sim 25$~Hz which we attribute to environmental mechanical noise, was present.
This is compensated by using the temporally fluctuating $\theta(t)=\theta(t=0) +\delta\theta(t)$ to displace the filter functions used in the data analysis. For  $|\delta \theta(t)| \ll \pi$,  the correction was effective. In the experiment although $|\delta \theta(t)|  \lesssim \pi/4$ was significant,  the correction systematically improved the strength of the sidebands by about $20\%$.
Agreement with theory remained excellent. The case (iii) shows some distortion relative to theory. 
We attribute this to frequency noise which distorted sideband profiles. The effect on peak amplitudes (at $\theta=0$ and $\theta=\pi/2$) is small as they still have amplitudes close to values predicted from  Eqs.\ref{rheterodyne} and \ref{rhett0}. Finally, although the experimental data
were in thermal regimes,
Fig.\ref{Fig2}(c) illustrates potential new signatures of quantum regimes which are possible in the presence of fast and accurate correlation mapping with future experiments with cryogenically cooled
set-ups and stronger optomechanical sideband cooling.

 Filtering, applied directly to optical signals, is effective in mitigating the
 deleterious effects of technical noise \cite{Filters}; here we show that a type of filtering of the {\em autocorrelation} can distil correlations
 which are otherwise entirely absent in the heterodyne PSD, even if it were noise-free.  Interest in detection of these correlations and related effects
is growing and motivating sophisticated experiments \cite{Sudhir2017,Purdy2017} which use multiple heterodyne fields \cite{Lehnert2017,Synodyne2016} or demodulation of the signal and cross-correlation spectra. In fact the spectra in Fig.\ref{Fig2}(i) correspond to the cross-correlation between red and blue sidebands $\langle i_h^*(\omega+\Omega)i_h(\omega-\Omega)\rangle+ \textrm{C.C.}$
(see \cite{Chang2011} for derivation).

 Fig.\ref{Fig3} illustrates a r-heterodyne spectrum that most closely approximates homodyne behavior by combining the correlations with the main
heterodyne peaks at the same frequency, yet retaining the split-peak heterodyne character which allows Stokes/anti-Stokes sideband asymmetry to be measured. We show also that with fast 2D
imaging (even in real-time), new types of quantum signatures are accessible
that are not evident in a single PSD; this is exemplified by Fig.\ref{Fig2}(a) panel (i).
These particular sidebands are still stronger than heterodyne for $\theta \simeq \pi/2$ and have the advantage that the pure correlation spectral sidebands exhibit no imprecision noise floor,
mitigating another cause of experimental uncertainty. 

In conclusion, our work opens up the possibility of sensing using  spectra which are of hybrid homodyne-heterodyne character,  that can simultaneously display (i) sideband asymmetries (a signatures that  mechanical motions are in the quantum regime) (ii) squeezing and quantum signatures of correlations between optical quadratures; and yet permitting ultrasensitive displacement detection with sidebands with combined amplitudes comparable to homodyne.  The r-heterodyne technique may also have further practical applications for optomechanical systems with non-stationary 
cavity dynamics \cite{MariEisert,PRL2016,Aranas2016}. 

\section*{Acknowledgements}
The authors acknowledge useful discussions with Erika Aranas, Andrew Higginbotham and Florian Marquardt. This project has received funding from the European Union's Horizon 2020 research and innovation programme under the Marie Sklodowska-Curie Grant Agreement No. 749709. The work received support from EPSRC grant EP/N031105.


\begin{thebibliography}{99}
\bibitem{homodyne} The measured quantity of interest in homodyne detection is the  symmetrised PSD
$\overline{S}_{Y_\theta Y_\theta}(\omega)=\frac{1}{2}[S_{Y_\theta Y_\theta}(\omega)+S_{Y_\theta Y_\theta}(-\omega)]$
\end{thebibliography}
 \end{document}